\documentclass[aps,prb,preprint,superscriptaddress,showpacs]{revtex4-1}
\usepackage{amsmath}%
\usepackage{amsfonts}%
\usepackage{amssymb}%
\usepackage[latin1]{inputenc}
\usepackage{graphicx}
\usepackage{epstopdf}
\usepackage{color}
\usepackage{ulem}
\usepackage[english]{babel}
\usepackage{natbib}

\begin{document}

\title{Impurity states in the magnetic topological insulator V:(Bi,Sb)$_2$Te$_3$}

\author{Thiago R. F. Peixoto}
\email[Corresponding address:]{thiago.peixoto@physik.uni-wuerzburg.de}
\affiliation{Experimentelle Physik VII and R\"{o}ntgen Research Center for Complex Materials (RCCM),
Fakult\"{a}t f\"{u}r Physik und Astronomie, Universit\"{a}t W\"{u}rzburg, Am Hubland, D-97074 W\"{u}rzburg, Germany}

\author{Hendrik Bentmann}
\affiliation{Experimentelle Physik VII and R\"{o}ntgen Research Center for Complex Materials (RCCM),
Fakult\"{a}t f\"{u}r Physik und Astronomie, Universit\"{a}t W\"{u}rzburg, Am Hubland, D-97074 W\"{u}rzburg, Germany}

\author{Steffen Schreyeck}
\affiliation{Experimentelle Physik III and R\"{o}ntgen Research Center for Complex Materials (RCCM),
Fakult\"{a}t f\"{u}r Physik und Astronomie, Universit\"{a}t W\"{u}rzburg, Am Hubland, D-97074 W\"{u}rzburg, Germany}

\author{Martin Winnerlein}
\affiliation{Experimentelle Physik III and R\"{o}ntgen Research Center for Complex Materials (RCCM),
Fakult\"{a}t f\"{u}r Physik und Astronomie, Universit\"{a}t W\"{u}rzburg, Am Hubland, D-97074 W\"{u}rzburg, Germany}

\author{Christoph Seibel}
\affiliation{Experimentelle Physik VII and R\"{o}ntgen Research Center for Complex Materials (RCCM),
Fakult\"{a}t f\"{u}r Physik und Astronomie, Universit\"{a}t W\"{u}rzburg, Am Hubland, D-97074 W\"{u}rzburg, Germany}

\author{Henriette Maa\ss}
\affiliation{Experimentelle Physik VII and R\"{o}ntgen Research Center for Complex Materials (RCCM),
Fakult\"{a}t f\"{u}r Physik und Astronomie, Universit\"{a}t W\"{u}rzburg, Am Hubland, D-97074 W\"{u}rzburg, Germany}

\author{Mohammed Al-Baidhani}
\affiliation{Experimentelle Physik VII and R\"{o}ntgen Research Center for Complex Materials (RCCM),
Fakult\"{a}t f\"{u}r Physik und Astronomie, Universit\"{a}t W\"{u}rzburg, Am Hubland, D-97074 W\"{u}rzburg, Germany}

\author{Katharina Treiber}
\affiliation{Experimentelle Physik VII and R\"{o}ntgen Research Center for Complex Materials (RCCM),
Fakult\"{a}t f\"{u}r Physik und Astronomie, Universit\"{a}t W\"{u}rzburg, Am Hubland, D-97074 W\"{u}rzburg, Germany}

\author{Sonja Schatz}
\affiliation{Experimentelle Physik VII and R\"{o}ntgen Research Center for Complex Materials (RCCM),
Fakult\"{a}t f\"{u}r Physik und Astronomie, Universit\"{a}t W\"{u}rzburg, Am Hubland, D-97074 W\"{u}rzburg, Germany}

\author{Stefan Grauer}
\affiliation{Experimentelle Physik III and R\"{o}ntgen Research Center for Complex Materials (RCCM),
Fakult\"{a}t f\"{u}r Physik und Astronomie, Universit\"{a}t W\"{u}rzburg, Am Hubland, D-97074 W\"{u}rzburg, Germany}

\author{Charles Gould}
\affiliation{Experimentelle Physik III and R\"{o}ntgen Research Center for Complex Materials (RCCM),
Fakult\"{a}t f\"{u}r Physik und Astronomie, Universit\"{a}t W\"{u}rzburg, Am Hubland, D-97074 W\"{u}rzburg, Germany}

\author{Karl Brunner}
\affiliation{Experimentelle Physik III and R\"{o}ntgen Research Center for Complex Materials (RCCM),
Fakult\"{a}t f\"{u}r Physik und Astronomie, Universit\"{a}t W\"{u}rzburg, Am Hubland, D-97074 W\"{u}rzburg, Germany}

\author{Arthur Ernst}
\affiliation{Max Planck Institute of Microstructure Physics, Weinberg 2, D-06120, Halle (Saale), Germany}
\affiliation{Wilhelm-Ostwald-Institut f\"{u}r Physikalische und Theoretische Chemie, Universit\"{a}t Leipzig, Linn\'{e}stra{\ss}e 2, D-04103 Leipzig, Germany}

\author{Laurens W. Molenkamp}
\affiliation{Experimentelle Physik III and R\"{o}ntgen Research Center for Complex Materials (RCCM),
Fakult\"{a}t f\"{u}r Physik und Astronomie, Universit\"{a}t W\"{u}rzburg, Am Hubland, D-97074 W\"{u}rzburg, Germany}

\author{Friedrich Reinert}
\affiliation{Experimentelle Physik VII and R\"{o}ntgen Research Center for Complex Materials (RCCM),
Fakult\"{a}t f\"{u}r Physik und Astronomie, Universit\"{a}t W\"{u}rzburg, Am Hubland, D-97074 W\"{u}rzburg, Germany}

\date{\today}

\begin{abstract}
The ferromagnetic topological insulator V:(Bi,Sb)$_2$Te$_3$ has been recently reported as a quantum anomalous Hall (QAH) system. Yet the microscopic origins of the QAH effect and the ferromagnetism remain unclear. One key aspect is the contribution of the V atoms to the electronic structure. Here the valence band of  V:(Bi,Sb)$_{2}$Te$_3$ thin films was probed in an element-specific way by resonant photoemission spectroscopy. The signature of the V $3d$ impurity band was extracted and exhibits a high density of states near Fermi level, in agreement with spin-polarized first-principles calculations. Our results indicate the occurrence of a ferromagnetic superexchange interaction mediated by the observed impurity band, contributing to the ferromagnetism in this system. 
\end{abstract}

\pacs{73.20.Hb; 75.50.Pp; 78.70.Dm; 79.60.Dp}

\keywords{magnetic topological insulators; transition metal impurities; quantum anomalous Hall effect; Van-Vleck ferromagnetism; Zener double exchange interaction; ferromagentic superexchange interaction; $d$-electron impurity band}

\maketitle

Magnetic topological insulators (MTI) are a novel class of materials that unite the existence of a topologically non-trivial electronic band structure with the long-range ferromagnetic order induced by magnetic impurities in the system \cite{Hasan2010,Yu2010,Jiang2015,Weng2015}. The bulk electronic structure of these materials features a non-trivial symmetry inversion at the bandgap due to strong spin-orbit coupling (SOC). This gives rise to helical edge or surface states within the fundamental bandgap, which are protected against backscattering by time-reversal symmetry. The ferromagnetic ground state breaks time-reversal symmetry, affecting the topological properties of the system. The inverted bulk band structure and the insulating ferromagnetic ground state are predicted to be the pre-requisites for the realization of the quantum anomalous Hall state (QAHS), a dissipationless spin-polarized quantized transport state in the absence of external magnetic fields \cite{Weng2015}. The QAHS in MTI offers promising prospects towards the advent of low energy-consumption electronic devices \cite{Hasan2010,He2013,Eschbach2015,Weng2015}, and therefore has been attracting wide attention in condensed matter physics. In fact, the QAHS has been recently reported on two MTI systems, namely V- and Cr-doped (Bi$_x$Sb$_{1-x}$)$_2$Te$_3$ epitaxial thin films, in ultra-low temperature ($<100$ mK) magnetotransport studies \cite{Chang2013,Chang2015,Jiang2015,Grauer2015}.  

In spite of these exciting developments, the mechanism driving the ferromagnetism in these systems remains under intense debate \cite{Larson2008,Yu2010,Vergniory2014,Lachman2015,Grauer2015,Li2015,Ye2015,Sessi2016}. Key aspects are how the $3d$ magnetic impurities are incorporated in the TI structure and how their presence affects the electronic structure. Thus, theoretical works have focused their efforts on the prediction of the impurities density of states (DOS) and the magnetic interactions in model MTI systems. First-principles calculations have predicted a metallic impurity band at the Fermi level ($E_F$) for different V-doped TIs, yet a fully gapped band structure for Cr-doped systems \cite{Larson2008,Yu2010,Zhang2012,*Zhang2013,Vergniory2014}. Experimentally, it turned out that both V- and Cr-doped (Bi$_x$Sb$_{1-x}$)$_2$Te$_3$ are ferromagnetic topological insulators at temperatures below 30 K, with rather robust magnetic properties for the former, and a weak magnetism for the latter \cite{Chang2015}. Clearly the $3d$-states of the transition metal impurities play a very important role on the magnetism in MTI, \textit{e.g.} by providing additional mechanisms of ferromagnetic coupling \cite{Ye2015,Dietl2000,Dietl2007,Belhadji2007,Dietl2010,Sato2010}, although yet not experimentally determined. 

Recently, Vergniory \textit{et al.} \cite{Vergniory2014} have shown that the magnetism picture in these systems may be more complex than the Van-Vleck mechanism so far discussed \cite{Yu2010,Li2015,Chang2013,Chang2015}. Their calculations showed that the exchange constants (magnitude and sign) strongly depend on the filling of the $d$-states, and that competing coupling mechanisms may simultaneously take place. The magnetic impurities couple via an effective exchange mediated by the $p$-orbitals of the chalcogen atoms of the host, in a fashion well described by the Zener model \cite{Zener1951} in the context of diluted magnetic semiconductors (DMS) \cite{Dietl2000,Dietl2007,Belhadji2007,Sato2010,Dietl2010}. The inverted orbital symmetry at the bandgap introduces very unique physical conditions, which pose new challenges to the understanding of the valence band magnetism in MTI. Therefore it is of great importance to know in detail the electronic structure at the valence band, and in particular to identify the nature of the electronic states near $E_F$ in these systems. In this Rapid Communication we exploit the element-specificity of resonant photoemission (resPES), X-ray photoemission (XPS) and absorption spectroscopy (XAS) to study the signature of V impurities in the electronic structure of V$_y$(Bi$_x$Sb$_{1-x}$)$_{2-y}$Te$_3$ epitaxial thin films. With the help of spin-polarized first-principles calculations, we are able to identify the impurity-band character of the V $3d$ states at the valence band. Here we show that these states fulfill the conditions for mediating a ferromagnetic superexchange interaction between magnetic impurities, thereby contributing to the onset of ferromagnetism and QAHS at low temperatures \cite{Yu2010,Vergniory2014,Chang2015}.   

The systems studied consist of 10 nm thick (Bi$_{x}$Sb$_{1-x}$)$_{2}$Te$_3$ epitaxial films, doped with V atoms, grown on a Si(111) substrate. The V$_y$(Bi$_{x}$Sb$_{1-x}$)$_{2-y}$Te$_3$ samples have a Bi concentration of $x=0.21$, and V concentrations of $y=0.1$ (2 at.\% ) and $y=0.2$ (4 at.\%). The MBE growth has been described elsewhere \cite{Grauer2015}. Undoped samples were used as reference in the experiments. After growth, the films were cooled to a temperature below 283 K and capped by a 200 nm thick amorphous Se layer, in order to protect the surface from contamination during sample transport. Prior to the measurements, the samples were decapped by thermal desorption at 400 K under a pressure of $1\times10^{-9}$ mbar. As reported previously \cite{Maass2014}, this procedure results in a significant Se/Te substitution at the surface. The XPS and XAS measurements were performed in the ASPHERE III endstation, using a Scienta R4000 electron analyzer, with an energy resolution better than 200 meV, at the XUV beamline P04 of PETRA III (DESY, Hamburg) \cite{Viefhaus2013}, at a base pressure of $3\times10^{-10}$ mbar. All measurements shown here were performed at room temperature. The Fermi edge calibration was performed on a polycrystalline metallic surface in electrical contact with the sample. First-principles calculations were performed using a fully relativistic Green function method \cite{Geilhufe2015} within density functional theory in a generalized gradient approximation \cite{Perdew1996}. V impurities in (Bi$_{x}$Sb$_{1-x}$)$_{2}$Te$_3$ were assumed to randomly substitute atoms in the cation layers, and described within a coherent potential approximation as it is implemented within the multiple scattering theory \cite{Soven1967,Gyorffy1972}. The calculations were performed for V concentrations from $y=0$ to 0.2, and a Bi concentration of $x=0.21$.

\begin{figure}[htbp]
\includegraphics[width=0.49\textwidth]{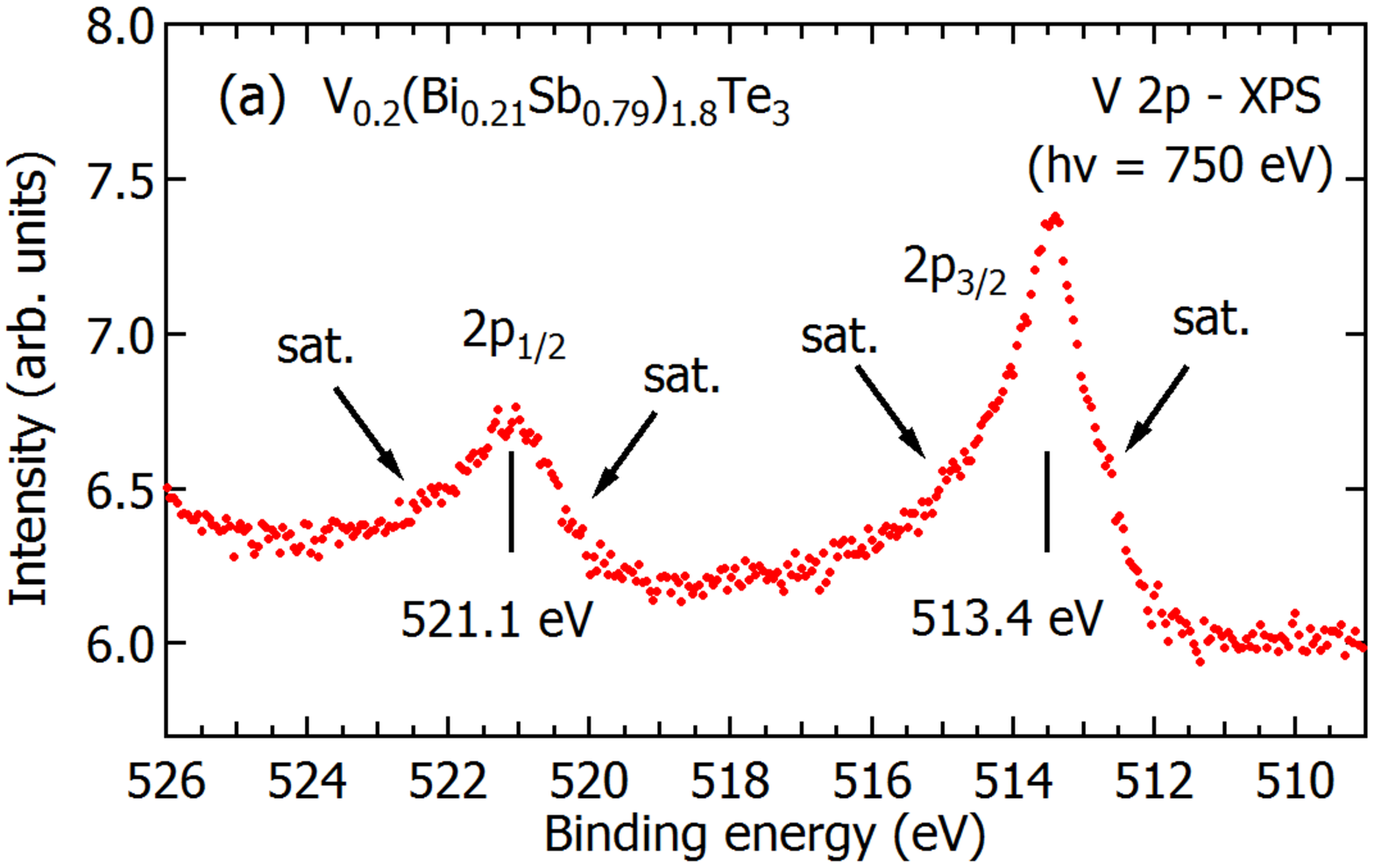}
\includegraphics[width=0.49\textwidth]{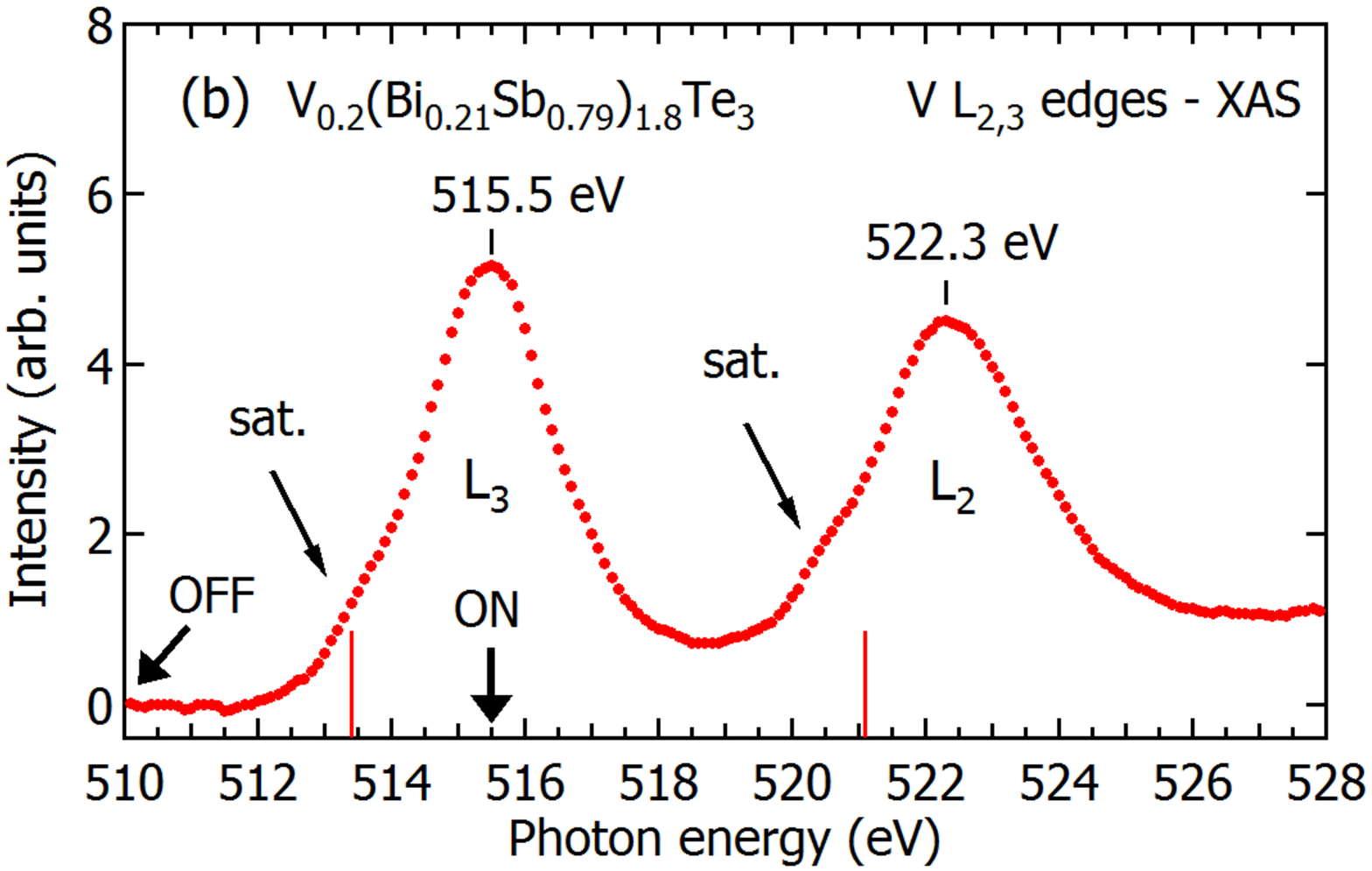} 
\caption{Representative spectra of (a) the V $2p$ core levels, measured at $h\nu=750$ eV, and (b) the V $L_{2,3}$ edges from a V$_{0.2}$(Bi$_{0.21}$Sb$_{0.79}$)$_{1.8}$Te$_3$ (4 at.\%) film. Satellite structures (labeled ``sat.'') characterize the V $2p$ and $L$-edge lineshapes in this system. The red lines mark the position of the V $2p$ core levels. The $L_3$-edge maximum is found 1.9 eV above the $2p_{3/2}$ peak. The labels ON and OFF mark the photon energies used in the resPES measurements.
\label{fig1}}
\end{figure}
In figure \ref{fig1}(a) we show a representative XPS spectrum of the V $2p$ core levels from a V$_{0.2}$(Bi$_{0.21}$Sb$_{0.79}$)$_{1.8}$Te$_3$ film, measured with $h\nu=750$ eV. The V $2p_{3/2}$ and $2p_{1/2}$ peaks are located at 513.4 eV and 521.1 eV, respectively, resulting in a spin-orbit splitting of $\Delta_{SO}=7.7$ eV. The chemical shift of about 1.3 eV towards higher binding energies ($E_B$) with respect to metallic V \cite{Kurmaev1998,Ilakovac2005} indicates an effectively reduced valence of the V atoms in the MTI structure. The peak lineshapes also exhibit satellite contributions in respect to pure V, most likely due to final-state effects \cite{Fink1985,Heide2008,Groot2005}. In addition, we characterized the V $L_{2,3}$ absorption edges by XAS. As shown in fig. \ref{fig1}(b), the positions of the $L_3$ and $L_2$ maxima occur at $h\nu=515.5$ eV and $h\nu=522.3$ eV, respectively, similar to metallic V \cite{Scherz2004,Ilakovac2005}. These are 1.9 and 1.2 eV above the position of the measured $2p$ XPS lines (marked by the red lines). The lineshapes, on the other hand, show: small satellite structures on the low energy side of the $L$-edges, deviation from the statistical $L_3/L_2$ intensity ratio of (2:1), and a reduction of the peak maxima distance of about 0.9 eV with respect to our XPS results. These observations have been earlier attributed to strong core-hole Coulomb and exchange interactions in the final state for early $3d$ transition metals \cite{Fink1985,Zaanen1985,Scherz2004,Ilakovac2005,Groot2005}. Similar results (XPS and XAS) were obtained for the 2 at.\% doped sample. Our XAS results also show good agreement with recent studies of the V $L$-edges in V-doped Sb$_2$Te$_3$ \cite{Li2015,Sessi2016}, and the scanning tunneling microscopy characterization of that ingot confirmed a dilute character of the V doping (0.75\%), \textit{i.e.} random distribution of V atoms substituting Sb, and no signs of clustering \cite{Sessi2016}. The XPS and XAS lineshapes may be strongly influenced by the chemical environment of the impurity atoms and the unoccupied states \cite{Fink1985,Zaanen1985,Kurmaev1998,Zimmermann1998,Scherz2004,Ilakovac2005,Groot2005,Heide2008}. The V $2p$ and $L_{2,3}$ lineshapes observed in V:(Bi,Sb)$_2$Te$_3$ resemble those from tetravalent charge-transfer V compounds, such as VO$_2$ \cite{Abbate1993,Kurmaev1998,Surnev2003,Zimmermann1998,Demeter2000,Surnev2003,Heide2008} and VS$_2$ \cite{Mulazzi2010}, where the V atoms are found on octahedral sites. Although we cannot exclude local inhomogeneities of the V distribution, our results are consistent with the scenario of V atoms substituting Sb(Bi), with a local octahedral coordination, as theoretically predicted \cite{Yu2010,Zhang2012,*Zhang2013,Vergniory2014} and experimentally observed in dilute V:Sb$_2$Te$_3$ \cite{Sessi2016}.

\begin{figure}[htbp]
\includegraphics[width=0.49\textwidth]{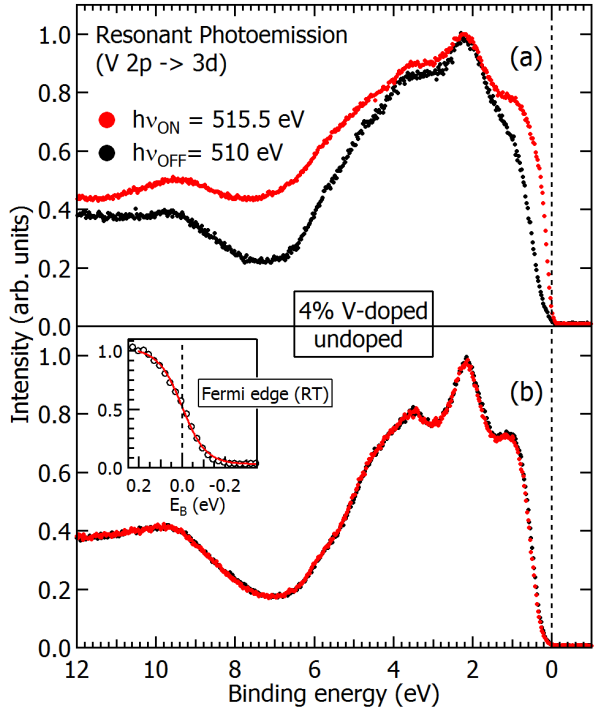} 
\caption{(a) Resonant photoemission spectra of the valence band of V$_{0.2}$(Bi$_{0.21}$Sb$_{0.79}$)$_{1.8}$Te$_3$, at the $L_3$ absorption edge ($2p \rightarrow 3d$ transition). The red and black curves were measured in on- ($h\nu_{ON}=515.5$ eV) and off-resonant ($h\nu_{OFF}=510$ eV) conditions, respectively. (b) Same set of measurements as in (a), now performed on an undoped (Bi$_{0.21}$Sb$_{0.79}$)$_{2}$Te$_3$ reference sample. The absence of V in the sample leads to no resonant enhancement of the valence band photoemission. The inset shows the Fermi edge calibration at room temperature.
\label{fig2}}
\end{figure}
On the basis of the measured V $L$-edges at the V$_{0.2}$(Bi$_{0.21}$Sb$_{0.79}$)$_{1.8}$Te$_3$ films, we performed resPES at the V $L_3$-edge ($h\nu_{ON}=515.5$ eV, ``on-resonance"), in order to highlight the contribution of the V $3d$-states in the valence band. In resPES, the valence band photoemission is enhanced by the resonant V $2p\rightarrow3d$ transition, \textit{i.e.} in an element-specific way \cite{Saisa1992,Martensson1997,Huefner2000}. For comparison, similar PES spectra were measured at an energy below the absorption edge ($h\nu_{OFF}=510$ eV, ``off-resonance"). In figure \ref{fig2}(a) we plot the on- and off-resonance spectra (red and black curves, respectively, as described above) measured on that sample, and normalized by the non-resonant high binding energy tail (above 13 eV). The clear difference observed between these two curves is due to the contribution of the V $2p\rightarrow3d$ transition to the PES in the on-resonance condition. Also a contribution from direct Auger decay is observed at higher binding energies ($E_B\geq4$ eV), as extensively reported by resPES studies at the $L$-edges of transition metals \cite{Saisa1992,Pervan1996,Weinelt1997,Martensson1997,Huefner2000,Ilakovac2005}. As a control, the same set of measurements on a (Bi$_{0.21}$Sb$_{0.79}$)$_{2}$Te$_3$ reference sample, containing no V, are shown in figure \ref{fig2}(b). As expected, no difference between the on- and off-resonance spectra is observed in this case. 

\begin{figure}[htbp]
\includegraphics[width=0.49\textwidth]{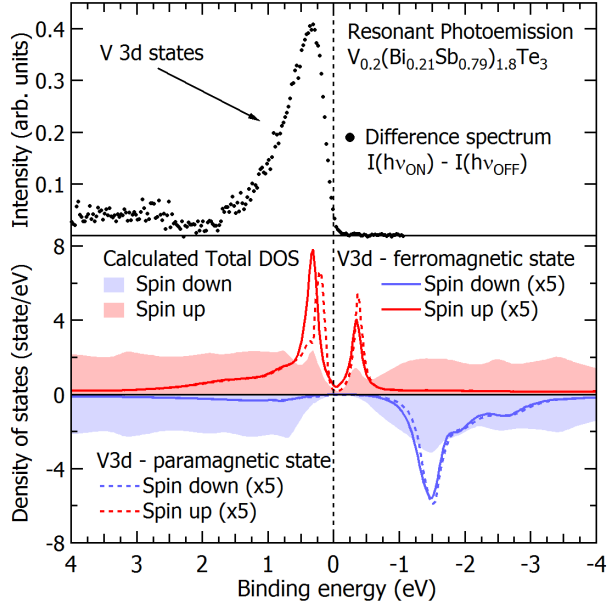}
\caption{V $3d$ impurity states at the valence band of V$_{0.2}$(Bi$_{0.21}$Sb$_{0.79}$)$_{1.8}$Te$_3$. Upper panel: resPES difference spectrum $I(h\nu_{ON}) - I(h\nu_{OFF})$ [from fig. \ref{fig2}(a)], representing the contribution of the V $3d$-states to the valence band. Lower panel: calculated spin-resolved DOS (shaded areas). The highlighted exchange-split bands (full lines and dashed lines) represent the spin-polarized V $3d$ impurity states (in the ferromagnetic and paramagnetic states, respectively). The peak in the resPES difference spectrum is attributed to the V spin-up band below $E_F$ due to their remarkable agreement.\label{fig3}}
\end{figure}

In order to extract the contribution of the V $3d$ states to the valence band, we plot the difference between the on- and off-resonance spectra from fig. \ref{fig2}(a) in fig. \ref{fig3} (upper panel). It shows a clear peak at roughly 0.3 eV below $E_F$, spreading down to $E_B\approx2$ eV, and with a finite intensity at $E_F$. This peak is attributed to the formation of an impurity band near the valence band maximum. To support this interpretation, we performed \textit{ab initio} calculations of the spin-polarized DOS for the V$_{0.2}$(Bi$_{0.21}$Sb$_{0.79}$)$_{1.8}$Te$_3$ system, and the results are shown in the lower panel of fig. \ref{fig3}. In this plot the majority- and minority-spin DOS are separately depicted as red and blue, respectively. The shaded areas represent the total DOS, \textit{i.e.} the contributions of the host and impurity states together, whereas the full lines represent the sole contributions of V $3d$-states (five times magnified) in the ferromagnetic state. The agreement between our resPES results and the theory is good. Our calculations predict an exchange-split ground state near the valence band maximum, with a calculated V magnetic moment of 2.57 $\mu_B$ and an exchange splitting of about 1.8 eV. The majority-spin states are split by the crystal field into a lower band, sharply peaked at 0.31 eV below $E_F$, and an upper band found at 0.35 eV above $E_F$. The minority-spin states are completely unoccupied, spanning from 0.35 to 4.00 eV above $E_F$, with a maximum at $E_B=-1.50$ eV. As in the experiment, the calculations also indicate a finite intensity at $E_F$, near the top of the lower majority-spin band. Calculations of the spin-polarized DOS were also performed in the paramagnetic state (dashed lines in fig. \ref{fig3}), within the disordered local moment method \cite{Oguchi1983,Gyorffy1985,Staunton1985}. They only show minor changes in the V $3d$ DOS and no changes in the calculated exchange splitting and atomic magnetic moment, in comparison to the ferromagnetic state. In the paramagnetic state, the occupied majority-spin state lies at 0.2 eV below $E_F$, while the unoccupied majority-spin state shifts about 15 meV away from $E_F$. Our calculations indicate that the V $3d$ states are impurity-like, and do not strongly hybridize with the TI bands (even for the concentration of 4 at.\%). This has been previously attributed to size mismatch between the V atom and the Sb(Bi) substituted atoms \cite{Larson2008,Zhang2012,*Zhang2013}. In addition to the substitutional disorder, the presence of interstitial V dopants in the crystalline structure and at the van der Waals gaps cannot be excluded \cite{Zhang2012,*Zhang2013}. The contribution of these atoms to the signature of the V $3d$ states could differ from that predicted by our calculations. This further source of disorder could lead to different local symmetries of the crystal-field splitting and possibly different $pd$ hybridization schemes that could influence the dispersion of the $3d$ states. These may be some of the reasons for the discrepancies observed between experiment and theory, such as the broader bandwidth and the slightly different lineshape of the experimental impurity band.

Our results on the V $3d$ DOS at the valence band of V:(Bi,Sb)$_2$Te$_3$ may have important implications to the mechanisms of magnetic coupling in this system. As formulated by Anderson \cite{Anderson1950} and summarized by the Kanamori-Goodenough rules \cite{Goodenough1955,Kanamori1959}, a ferromagnetic superexchange interaction may arise between cations with a less-than-half-filled $d$-shell, subject to the crystal field of surrounding anions. This interaction is indirectly mediated by the overlap between $p$ and $d$ states of neighboring anions and cations, respectively, which allows a spin-dependent electron transfer \cite{Anderson1950,Kanamori1959}. One well established case where the ferromagnetism is stabilized by a ferromagnetic superexchange interaction is the DMS V:GaAs \cite{Belhadji2007,Sato2010}. In that case, $E_F$ lies in the gap between the lower and upper crystal-field-split majority-spin bands, coupling neighboring impurities ferromagnetically \cite{Belhadji2007,Sato2010}. With the increase of V concentration, the overlap between states from neighboring impurities should increase, broadening and pushing the lower majority band to higher $E_B$, and stabilizing the ferromagnetic coupling \cite{Belhadji2007}. Our calculated V $3d$ DOS exhibits a similar structure as in V:GaAs, and did reproduce these trends as a function of V concentration (not shown here), indicating a finite DOS at $E_F$ for the 4 at.\% ($y=0.2$) V concentration, as observed in our resPES experiments. On the other hand, $E_F$ being located in the gap between the lower and upper majority-spin bands and the finite DOS at $E_F$ are also suitable conditions for the occurrence of a ferromagnetic Zener-type double exchange interaction \cite{Zener1951a,Anderson1955,Sato2010}. This interaction relies on the strength of the $pd$ hybridization of the states at $E_F$, and has a maximum when $E_F$ lies in the middle of the impurity band, \textit{i.e.} when only bonding states are occupied \cite{Sato2010}. In the general case, these two magnetic coupling mechanisms are simultaneously present to some extent and cannot be easily decoupled. If, however, the broadening causes the top of the occupied band to reach $E_F$, affecting the $d$-shell filling, the superexchange interaction decays in favor of double exchange interaction \cite{Belhadji2007,Sato2010}. 

In the present case, we find the maximum of the V impurity band well below $E_F$, close to a scenario dominated by ferromagnetic superexchange interaction. Nevertheless, both our calculations and experiments indicate a finite V $3d$ DOS at $E_F$, suggesting an additional relevance of double exchange interaction. Thus, the V $3d$ DOS in the valence band of V:(Bi,Sb)$_2$Te$_3$ fulfills the conditions for the coexistence of double exchange and ferromagnetic superexchange interactions, both of them being very sensitive to small changes of the V DOS at $E_F$. At low temperatures, small variations of the $3d$ impurity band affecting the $d$-shell filling and the bandwidth at $E_F$, \textit{e.g.} localization of the $3d$ impurity band, may strongly favor the ferromagnetic superexchange, and influence the onset of ferromagnetism and QAHS recently reported in these systems \cite{Yu2010,Ye2015,Chang2015,Grauer2015}.
 
In summary, we have determined the signature of the V $3d$-electron impurity states at the valence band of the two-dimensional MTI V:(Bi,Sb)$_2$Te$_3$ by resonant photoemission. The occupied $3d$ impurity band is localized near the valence band maximum, in good agreement with \textit{ab initio} calculations of the spin-polarized DOS. The structure of the calculated V $3d$ impurity states and the weak $pd$ hybridization with the host favor the mediation of a ferromagnetic superexchange interaction, based on the hybridization between spin-polarized $3d$ impurity bands of neighboring V atoms. This interaction may possibly coexist with a double exchange interaction, offering additional mechanisms for establishing an insulating ferromagnetic ground state at low temperatures, along with the Van-Vleck mechanism \cite{Yu2010,Chang2015,Li2015,Ye2015}. Finally, we hope that this work stimulates further investigations of the intricate details of the DOS, \textit{e.g.} temperature dependence, effect of the host doping stoichiometry, and the valence band magnetism in the vicinity of the coupled quantum and magnetic phase transitions observed in MTI.

\begin{acknowledgments} 
Parts of this research were carried out at the PETRA III light source at DESY, a member of the Helmholtz Association (HGF). We would like to thank Dr. Jens Viefhaus, J\"{o}rn Seltmann and Frank Scholz for the assistance in using beamline P04. We would also like to express our gratitude to Dr. Matthias Kall\"{a}ne, from the Christian-Albrechts-Universit\"{a}t zu Kiel, for their assistance in using the ASPHERE III endstation. We are also thankful to Stephan Borsitzki for his assistance in the experiments during the beamtime. We gratefully acknowledge the financial support of the EU ERC-AG Program (project 3-TOP) and the DFG through the Priority Programm 1666 ``Topological Isolators'' and the SFB 1170 ``ToCoTronics'' (projects A01, B01 and B02).
\end{acknowledgments} 

\bibliography{Refs_BiSbV2Te3}

\end{document}